\documentclass[11pt,draftcls,onecolumn]{IEEEtran}

\usepackage{color}
\usepackage{amsmath,amssymb,amsthm}
\usepackage{pgfplots}
\usepackage{psfrag}
\usepackage{mathrsfs}
\usepackage{mathtools}

\usepackage{footnote}
\usepackage{amsfonts}

\newcommand{\pp}[1]{\ensuremath{\mathbb P\left\{#1\right\}}}
\newcommand{\snr}{{\sf SNR}}
\newcommand{\fh}{\hat{f}}
\newcommand{\Cs}{\mathscr{C}}
\newcommand{\Rs}{\mathscr{R}}

\let\oldbrace\{
\def\{{\oldbrace\kern0.5pt}

\newcommand{\Cc}{\mathcal{C}}

\newcommand{\Rc}{\mathcal{R}}

\newcommand{\Xc}{\mathcal{X}}
\newcommand{\Yc}{\mathcal{Y}}

\newcommand{\Xh}{{\hat{X}}}

\newcommand{\xh}{{\hat{x}}}

\def\e{\epsilon}

\let\P\relax
\DeclareMathOperator\P{\sf P}

\DeclareMathOperator\C{C}

\newcommand{\sfrac}[2]{\mbox{\small$\displaystyle\frac{#1}{#2}$}}

\usepackage{tikz,tkz-graph}
\usepackage[mode=buildnew]{standalone}
\usetikzlibrary{matrix,positioning,fit,calc,shapes}
\pgfplotsset{compat=newest}

\newtheorem{theorem}{Theorem}
\newtheorem{proposition}{Proposition}
\newtheorem{definition}{Definition}
\newtheorem{corollary}{Corollary}
\newtheorem{remark}{Remark}

\newtheorem{example}{Example}

\ifCLASSOPTIONcompsoc \usepackage[caption=false,font=normalsize,labelfon t=sf,textfont=sf]{subfig} \else \usepackage[caption=false,font=footnotesize]{subfi
g} \fi

\begin{document}

\sloppy
\IEEEoverridecommandlockouts
\title{Communication versus Computation: Duality for multiple access channels and source coding}

%
%
%


\author{Jingge Zhu, Sung Hoon Lim, and Michael Gastpar
\thanks{This paper was presented in part at the 2017 IEEE Information Theory and Applications Workshop.}
\thanks{J. Zhu is with the Department of Electrical Engineering and Computer Science, University of California, Berkeley, 94720 CA, USA (e-mail: jingge.zhu@berkeley.edu).}
\thanks{S.~H.~Lim is with the Korea Institute of Ocean Science and Technology, Ansan, Gyeonggi-do, Korea (e-mail: shlim@kiost.ac.kr).}%
\thanks{Michael Gastpar is with the School of Computer and Communication Sciences, Ecole Polytechnique F\'ed\'erale, 1015 Lausanne, Switzerland
 (e-mail: michael.gastpar@epfl.ch).}
}



\maketitle

\begin{abstract}
Computation codes in network information theory are designed for the scenarios where the decoder is \textit{not} interested in recovering the information sources themselves, but only a  function thereof. K\"orner and Marton showed for distributed source coding that such function decoding can be achieved more efficiently than decoding the full information sources. Compute-and-forward has shown that function decoding, in combination with network coding ideas, is a useful building block for end-to-end communication.  In both cases, good computation codes are the key component in the coding schemes. In this work, we expose the fact that good computation codes could undermine the capability of the codes for recovering the information sources individually, \textit{e.g.}, for the purpose of multiple access and distributed source coding.  Particularly, we establish  duality results between the codes which are good for computation and the codes which are good for multiple access or distributed compression.
\end{abstract}

\begin{IEEEkeywords}
Function computation, code duality, multiple access channel, compute--forward, multi-terminal source coding, structured code.
\end{IEEEkeywords}

\section{Introduction}
\label{sec:intro}

\begin{figure}[!hbt]
\begin{center}
\footnotesize
\psfrag{e1}[c]{Encoder 1}
\psfrag{e2}[c]{Encoder 2}
\psfrag{d1}[c]{Decoder 1}
\psfrag{d2}[c]{Decoder 2}
\psfrag{m1}[c]{$M_1$}
\psfrag{m2}[c]{$M_2$}
\psfrag{x1}[c]{$X^n_1$}
\psfrag{x2}[c]{$X^n_2$}
\psfrag{y1}[c]{$Y^n_1$}
\psfrag{y2}[c]{$Y^n_2$}
\psfrag{mh1}[c]{$f_1(M_1, M_2)$}
\psfrag{mh2}[c]{$(M_1, M_2)$}
\psfrag{p1}[c]{$p_1(y_1|x_1, x_2)$}
\psfrag{p2}[c]{$p_2(y_2|x_1, x_2)$}
\includegraphics[scale=1]{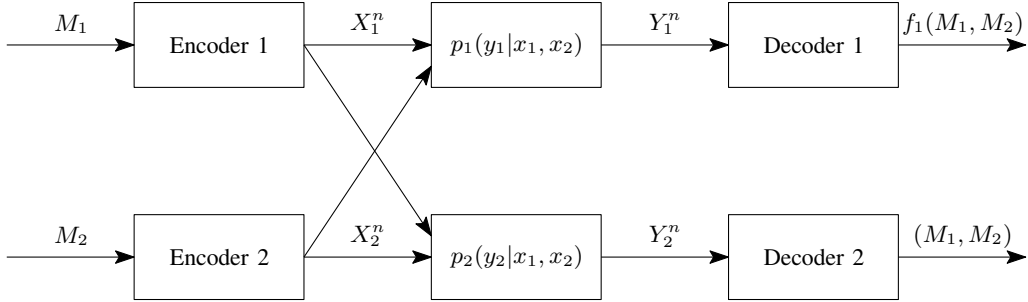}
\caption{Two-sender two-receiver network with channel distribution $W_1(y_1|x_1, x_2)W_2(y_2|x_1, x_2)$. Decoder 1 wishes to recover the sum of the codewords while Decoder 2 wishes to recover both messages.}
\label{fig:compoundMAC}
\end{center}
\end{figure}

To set the stage for the results and discussion presented in this paper, it is instructive to consider the two-sender two-receiver  memoryless network  illustrated in Fig.~\ref{fig:compoundMAC}. Specifically, this network consists of two multiple access channels that we allow to be different in general, characterized by their respective conditional probability distributions. The fundamental tension appearing in this network concerns the {\it decoders:}  Decoder 1 wishes only to recover a function $f_1(M_1, M_2)$ of the original messages.
By contrast, Decoder 2 is a regular multiple access decoder, wishing to recover both of the original messages. As illustrated in the figure, the tension arises because the two encoders must use one and the same code to serve both decoders.

In a memoryless Gaussian network where $X_1,X_2\in\mathbb R$, decoding the (element-wise) sum of the codewords $f_1(M_1, M_2)=x_1^n(M_1)+x^n_2(M_2)$ is often of particular interest. The computation problem associated with Decoder 1 is a basic building block for many  complex communication networks, including the well-known two-way relay channel \cite{wilson_joint_2010} \cite{Nam_etal_2010}, and general multi-layer relay networks \cite{NazerGastpar_2011}.  The \textit{computation} aspect of these schemes is important, sometimes even imperative in  multi-user communication networks. Results from network coding \cite{li_linear_2003} \cite{ahlswede_network_2000}, physical network coding \cite{zhang_hot_2006}, and the compute--forward  scheme \cite{NazerGastpar_2011} have all shown that computing certain functions of codewords within a communication network is vital to the overall coding strategy, and their performance cannot be achieved otherwise.  Previous studies have all suggested that good computation codes should possess some algebraic structures. For example, nested lattice codes are used in the Gaussian two-way relay channel and more generally in the  compute-and-forward scheme. In this case, the linear structure of the codes is the key to the coding scheme, due to the fact that multiple codeword pairs result in the same sum codeword, thus minimizing the number of competing sum codewords upon decoding.  

However, it turns out that this algebraic structure could be ``harmful", if the codes are used for the purpose of multiple-access. Roughly speaking, if the channel has a ``similar'' algebraic structure (looking at Fig.~\ref{fig:compoundMAC}, this would be the case if  $Y_2=X_1+X_2$), then the fact that multiple codeword pairs result in the same sum codeword (channel output)  makes it impossible for the individual messages to be recovered reliably.

In this paper, we show that there exists a fundamental conflict between codes for efficient computation and multiple access if the channel is matched with the algebraic structure of the function to be computed. One contribution of the paper is to give a precise statement of this phenomenon, showing a duality between the codes used for communication and the codes used for computation on the two-user multiple access channel (MAC). We show that codes which are ``good'' for computing certain functions over a multiple access channel will inevitably lose their capability to enable multiple access, and vice versa.  


Similar phenomena are observed in distributed source coding settings. We find that there exists a conflict between ``good" codes for computation and codes for reliable compression, as in the channel coding case.  In particular, we classify some fundamental conditions in which ``good'' computation codes cannot be used for recovering the sources separately. 

The paper is organized as follows. Beginning with the next section, we state the multiple access and computation duality results and provide the proofs of our theorems. 
In Section~\ref{sec:dsc-duality}, duality for computation and distributed source coding is given with some discussions and the proofs of the theorems. In Section~\ref{sec:GMAC}, we specialize the duality results for the Gaussian MAC. Finally, in Section~\ref{sec:discussions} we give some concluding remarks. Throughout the note, we will use $[n]$ to denote the set of integers $\{1,2,\ldots, n\}$ for some $n\in\mathbb Z_+$.

\section{Multiple Access and Computation Duality}\label{sec:mac-duality}

A two-user discrete memoryless multiple access channel (MAC) $(\mathcal X_1 \times \mathcal X_2, \mathcal Y, W(y|x_1,x_2))$ consists of three finite sets $\mathcal X_1,\mathcal X_2,\mathcal Y$, denoting the input alphabets and the output alphabets, respectively,   and a collection of conditional probability mass functions (pmf) $W(y|x_1,x_2)$. A formal definition of multiple access codes is given as follows.

\begin{definition}[multiple access codes]\label{def:ma_code}
A $(2^{nR_1}, 2^{nR_2}, n)$ multiple access code\footnote{or simply \textit{multiple access code}, when the parameters are clear from the context.} for a MAC consists of
\begin{itemize}
\item two message sets $[2^{nR_k}]$, $k=1,2,$
\item two encoders, where each encoder maps each message $m_k\in[2^{nR_k}]$ to a sequence $x_k^n(m_k)\in\mathcal X^n$ {\em bijectively},
\item a decoder that maps an estimated pair $(\hat x_1^n, \hat x_2^n)$ to each received sequence $y^n$.
\end{itemize}
Each message $M_k$, $k=1,2,$ is assumed to be chosen independently and uniformly from $[2^{nR_k}]$. The average probability of error for multiple access is defined as
\begin{align}
P_\e^{(n)}=\P\{(X_1^n, X_2^n) \neq (\Xh_1^n, \Xh^n_2)\}, \label{eq:pe-mac}
\end{align}
where $X_k^n=x^n_k(M_k)$.
We say a rate pair $(R_1, R_2)$ is {\em achievable for multiple access} if there exists a sequence of $(2^{nR_1}, 2^{nR_2}, n)$ multiple access codes such that $\lim_{n\to\infty}P_\e^{(n)}=0$. 
\end{definition}

The classical capacity results of the multiple access channel (see e.g. \cite{cover_elements_2006}) shows that for the MAC given by $W(y|x_1,x_2)$, there exists a sequence of $(2^{nR_1}, 2^{nR_2}, n)$ multiple access codes for any rate pair $(R_1,R_2)\in \Cs_\text{MAC}$ where $\Cs_\text{MAC}$ is the set of rate pairs $(R_1, R_2)$ such that
\begin{align*}
R_1 &< I(X_1;Y|X_2,Q)\\
R_2 &< I(X_2;Y|X_1,Q)\\
R_1+R_2 &< I(X_1,X_2;Y|Q).
\end{align*}
for some pmf $p(q)p(x_1|q)p(x_2|q)$.

The following definition formalizes the concept of computation codes used in this paper.
\begin{definition}[Computation codes for the MAC]\label{def:cp_code}
A $(2^{nR_1}, 2^{nR_2}, n, f)$ {\em computation code}\footnote{or simply \textit{computation code}, when the parameters are clear from the context.} for a MAC consists of two messages sets and two encoders defined as in Definition~\ref{def:ma_code} and
\begin{itemize}
\item a function $f:\mathcal X_1\times\mathcal X_2\mapsto \mathcal F$ for some image $\mathcal F$,
\item a decoder that maps an estimated function value $\fh^n\in\mathcal X^n$ to each received sequence $y^n$.
\end{itemize}
The message $M_k$, $k=1,2,$ is assumed to be chosen independently and uniformly from $[2^{nR_k}]$. The average probability of error for computation is defined as
\begin{align}\label{eq:pe-cp}
P_\e^{(n)}=\P\{F^n\neq \hat{F}^n\},
\end{align}
where $F^n=(f(X_{11}, X_{21}), \ldots, f(X_{1n}, X_{2n}) )$ denotes an element-wise application of the function $f$ on the pair $(X_1^n, X_2^n)$. We say a rate pair $(R_1, R_2)$ is {\em achievable for computation} if there exists a sequence of $(2^{nR_1}, 2^{nR_2}, n)$ computation codes such that $\lim_{n\to\infty}P_\e^{(n)}=0$.
\end{definition}

We note that since the function $f(X_1^n, X_2^n)$ can be computed directly at the receiver if the individual codewords $(X_1^n, X_2^n)$ are known, a $(2^{nR_1}, 2^{nR_2}, n)$ multiple access code for a MAC is readily a $(2^{nR_1}, 2^{nR_2}, n, f)$ computation code over the same channel for \textit{any} function $f$. More interesting are the computation codes with rates outside the MAC capacity region, i.e. $(R_1,R_2)\notin \Cs_\text{MAC}$. We refer to such codes as {\em good} computation codes for this channel. A formal definition of good computation codes is given as follows.

\begin{definition}[Good computation codes]\label{def:good_computation_codes}
Consider a sequence of $(2^{nR_1}, 2^{nR_2},n,f)$-computation codes for a MAC given by $W(y|x_1,x_2)$. We say that they are {\em good computation codes} for the MAC,  if $(R_1,R_2)$ is achievable for computation, and 
\begin{align*}
R_1+R_2> \max_{p(x_1)p(x_2)}I(X_1,X_2;Y)
\end{align*}
namely, the sum-rate of the two codes is larger than the sum capacity of the MAC.
\end{definition}


The  multiple access or computation capability of codes over a channel depends heavily on  the structure of the channel. To this end, we give the following definition of a multiple access channel.

\begin{definition}[$g$-MAC]
Given a function $g:\mathcal X_1\times \mathcal X_2\mapsto \mathcal F$ for some set $\mathcal F$, we say that a multiple access channel described by $W(y|x_1,x_2)$ is a $g$-MAC if the following Markov chain holds
\begin{align*}
(X_1,X_2)-g(X_1,X_2)-Y.
\end{align*}
\label{def:fMAC}
\end{definition}
For example, the Gaussian MAC 
\begin{align}\label{eq:gaussian-mac}
Y=x_1+x_2+Z
\end{align}
where $Z\sim\mathcal N(0, 1)$ is a $g$-MAC with $g(x_1,x_2):=x_1+x_2$.

\subsection{Main results}

In this subsection we show that for any sequence of codes, there is an intrinsic tension between their capability for computation and their capability for multiple access. Some similar phenomena have already been observed in \cite{lim_joint_2016} \cite{zhu_allerton_2016}. Here we make some precise statements.

\begin{theorem}[MAC Duality 1]
Consider a two-sender two-receiver memoryless channel in Fig.~\ref{fig:compoundMAC}. Assume that the multiple access channels are given by the conditional probability distributions $W_1(y_1|x_1,x_2)$ and $W_2(y_2|x_1,x_2)$, where the channel $W_2$ is a $g$-MAC. Further assume that a sequence of codes $(\Cc_1^{(n)},\Cc_2^{(n)})$ is good for computing the function $f$ over  $W_1$, namely the sum rate of the codes satisfies $R_1+R_2 >\max_{p(x_1)p(x_2)} I(X_1, X_2; Y_1)$. Then this sequence of codes cannot be used as multiple access codes for the channel $W_2$ (i.e., the receiver cannot decode both codewords correctly), if it holds that 
\begin{align}
H(g(X_1^n,X_2^n))\leq H(f(X_1^n,X_2^n)) \text{ as }n\rightarrow\infty
\label{eq:condition}
\end{align}
where the functions $f$ and $g$ are applied element-wise to the random vector pair $(X_1^n,X_2^n)$ induced by the codebooks $(\Cc_1^{(n)},\Cc_2^{(n)})$.
\label{thm:duality_1}
\end{theorem}

\begin{remark}
More precisely, we will show in the proof that the capacity region of the two-sender two-receiver network is bounded by 
\begin{align}
R_1+R_2 \le \max_{p(x_1)p(x_2)}I(X_1, X_2; Y_1). \label{eq:sumrate_bnd}
\end{align}
Notice that though decoder 2 is required to recover both messages separately, the capacity of the network is bounded by \eqref{eq:sumrate_bnd} which does not depend on $W_2$.
\end{remark}

\begin{remark}\label{rmk:GaussianvsDMC}
To avoid confusion, we recall that $X_k^n=x^n_k(M_k), k=1, 2$ are always discrete random variables (for both discrete memoryless and continuous memoryless channels), as the randomness is only induced from the random choice of the codeword from the given codebooks. More precisely, we have
\begin{align*}
\pp{X_k^n=x_k^n}=
\begin{cases}
\frac{1}{2^{nR_k}} &\quad\text{ if }x_k^n\in\mathcal C_k^{(n)}\\
0 &\quad \text{ otherwise.}
\end{cases}
\end{align*}
for $k=1,2$.
In Section~\ref{sec:GMAC} we give some specialized results explicitly for the Gaussian multiple access channel. 
\end{remark}

\begin{remark}
 We also point out that the entropy of the function $f(X_1^n,X_2^n), g(X_1^n,X_2^n)$ depends on the structure of the codebooks $\mathcal C_1^{(n)}, \mathcal C_2^{(n)}$, and it is in general difficult to verify the condition in (\ref{eq:condition}). Nevertheless an interesting special case is  $f=g$ where this condition  is trivially satisfied.
\end{remark}


The following theorem gives a complementary result.
\begin{theorem}[MAC Duality 2]\label{thm:duality_2}
Consider two memoryless multiple access channels given by the conditional probability distributions $W_1(y_1|x_1,x_2)$ and $W_2(y_2|x_1,x_2)$, where the channel $W_2$ is a $g$-MAC. If $(\Cc_1^{(n)},\Cc_2^{(n)})$ is a sequence of multiple access codes for the channel $W_2$, then it cannot be a good computation code w.r.t. the function $f$ over $W_1$, if it holds that
\begin{align*}
H(g(X_1^n,X_2^n))\leq H(f(X_1^n,X_2^n)) \text{ as }n\rightarrow \infty.
\end{align*}
\end{theorem}

Before presenting the proofs of the above tow theorems, we  give a few examples to illustrate the results. 

\subsection{Examples}

\begin{example}
If two codes $\mathcal C_1^{(n)}, \mathcal C_2^{(n)}\subseteq\mathbb R^n$ are good for computing the sum $x_1^n+x_2^n$ over the Gaussian MAC
\begin{align*}
Y_1^n=x_1^n+x_2^n+\tilde Z^n,
\end{align*}
then they cannot be used for multiple access for the Gaussian MAC 
\begin{align*}
Y_2^n=x_1^n+x_2^n+Z^n
\end{align*}
where $\tilde Z^n, Z^n$ are two i.i.d. Gaussian noise sequences with arbitrary variances. The result holds according to Theorem \ref{thm:duality_1} by choosing
\begin{align*}
f(x_1,x_2)=g(x_1,x_2)=x_1+x_2
\end{align*}
We will discuss more about this example in Section \ref{sec:GMAC}.
\end{example}

\begin{example}
If two codes $\mathcal C_1^{(n)}, \mathcal C_2^{(n)}\subseteq\mathbb R^n$ are good for computing the sum $a_1x_1^n+a_2x_2^n$ over \textit{any} Gaussian MAC, 
then they cannot be used for multiple access for the Gaussian MAC 
\begin{align*}
Y^n=a_1x_1^n+a_2x_2^n+Z^n.
\end{align*}
The result holds according to Theorem \ref{thm:duality_1} by choosing
\begin{align*}
f(x_1,x_2)=g(x_1,x_2)=a_1x_1+a_2x_2
\end{align*}
with arbitrary $a_1,a_2\in\mathbb R$.
\end{example}

\begin{example}
If two codes $\mathcal C_1^{(n)}, \mathcal C_2^{(n)}\subseteq \{0,1\}^n$ are good for computing the sum $x_1^n + x_2^n$ over \textit{any} MAC,
then they cannot be used for multiple access for the  MAC 
\begin{align*}
Y^n=x_1^n\cdot x_2^n.
\end{align*}
Here $x_1^n+x_2^n \in\{0,1,2\}^n$ and $x_1^n\cdot x_2^n\in\{0,1\}^n$ represent the element-wise  sum and product of $x^n_1$ and $x^n_2$ in $\mathbb R^n$, respectively. The result holds according to Theorem \ref{thm:duality_1} by choosing
\begin{align*}
f(x_1,x_2)&=x_1+x_2\\
g(x_1,x_2)&=x_1\cdot x_2.
\end{align*}
It is easy to see that $H(X_1^n+X_2^n)\geq H(X_1^n\cdot X_2^n)$ in  this case.

\end{example}

\begin{example}
If two codes $\mathcal C_1^{(n)}, \mathcal C_2^{(n)}\subseteq \{0,1\}^n$ are good for computing the element-wise product $x_1^n\cdot x_2^n$ over \textit{any} MAC,
then they cannot be used for multiple access for the  MAC 
\begin{align*}
Y^n=x_1^n\cdot x_2^n+Z^n
\end{align*}
where $Z^n\in\{0,1\}^n$ denotes an i.i.d. noise sequence independent of the channel inputs. The result holds according to Theorem \ref{thm:duality_1} by choosing
\begin{align*}
f(x_1,x_2)=g(x_1,x_2)=x_1\cdot x_2
\end{align*}
and the fact that $Y^n = x_1^n\cdot x_2^n+Z^n$ is a $g$-MAC.
\end{example}

A summary of the above examples is given in Table \ref{tab:examples}.

\begin{table}[!hb]
\begin{center}
  \begin{tabular}{ l |  c |  r }
    \hline
   alpabet $\mathcal X$& If $(\mathcal C_1^{(n)}, \mathcal C_2^{(n)})$ are good for computing $f(x_1^n,x_2^n)$ over \textit{any} MAC  &  cannot be used for multiple access over \\   \hline  
    \\[-0.8em]
  $x_1,x_2\in\mathcal X=\mathbb R$ & $f(x_1^n,x_2^n)=x_1^n+x_2^n$  & $Y^n=x_1^n+x_2^n+ Z^n$\\
  $x_1,x_2\in \mathcal X=\mathbb R$ & $f(x_1^n,x_2^n)=a_1x_1^n+a_2x_2^n$  &$Y^n=a_1x_1^n+a_2x_2^n+ Z^n$ \\
  $x_1,x_2\in \mathcal X=\{0,1\}$   & $f(x_1^n,x_2^n)=x_1^n+ x_2^n$   &$Y^n=x_1^n\cdot x_2^n$\\
    $x_1,x_2\in \mathcal X=\{0,1\}$   & $f(x_1^n,x_2^n)=x_1^n\cdot x_2^n$   &$Y^n=x_1^n\cdot x_2^n+Z^n$\\
\hline
  \end{tabular}
      \caption{A summary of the examples.}
            \label{tab:examples}
      \end{center}

\end{table}

\subsection{Proofs}

\begin{IEEEproof}[Proof of Theorem \ref{thm:duality_1}]
We consider two multiple access channels $W_1, W_2$ as described in the theorem. We assume temporarily that the pair of codes $(\mathcal C_1^{(n)}, \mathcal C_2^{(n)})$ are used for computation over $W_1$, and used for multiple access over $W_2$. In other words, the function $f(X_1^n,X_2^n)$ can be  decoded  reliably using $Y_1^n$, and the pair $(X_1^n,X_2^n)$ can be decoded reliably using $Y_2^n$. 
Under this assumption, an upper bound on the sum-rate $R_1+R_2$ can be derived as follows:
\begin{align*}
n(R_1+R_2) &=H(X_1^n,X_2^n)\\
&=I(X_1^n, X_2^n; Y^n_2)+H(X_1^n,X_2^n|Y_2^n)\\
&\stackrel{(a)}{\leq} I(X_1^n, X_2^n; Y^n_2)+n\epsilon_n\\
&\stackrel{(b)}{=} I(g(X_1^n, X_2^n);Y_2^n)+n\epsilon_n\\
&\leq H(g(X_1^n,X_2^n))+n\epsilon_n\\
&=H(f(X_1^n,X_2^n))+H(g(X_1^n,X_2^n))-H(f(X_1^n,X_2^n))+n\epsilon\\
&=I(Y_1^n;f(X_1^n,X_2^n))+H(f(X_1^n,X_2^n)|Y_1^n)+H(g(X_1^n,X_2^n))-H(f(X_1^n,X_2^n))+n\epsilon_n\\
&\stackrel{(c)}{\leq}  I(Y_1^n;f(X_1^n,X_2^n))+H(g(X_1^n,X_2^n))-H(f(X_1^n,X_2^n))+2n\epsilon_n\\
&\leq I(Y_1^n,X_1^n,X_2^n)+H(g(X_1^n,X_2^n))-H(f(X_1^n,X_2^n))+2n\epsilon_n\\
&\leq \sum_{i=1}^nI(Y_{1i};X_{1i},X_{2i})+H(g(X_1^n,X_2^n))-H(f(X_1^n,X_2^n))+2n\epsilon_n,
\end{align*}
Step $(a)$ follows from Fano's inequality under the assumption that $X_1^n, X_2^n$ can be decoded over $W_2$. Step $(b)$ holds since $W_2$ is a $g$-MAC, which implies the Markov chain $(X_1,X_2)-g(X_1,X_2)-Y$ for any choice of codes, hence $I(g(X_1^n, X_2^n);Y_2^n)=I(X_1^n,X_2^n;Y_2^n)$. Step $(c)$ follows from Fano's inequality under the assumption that $f(X_1^n,X_2^n)$ can be decoded over $W_1$.  The last step is due to the memoryless property of $W_1$.  Since $\epsilon_n\rightarrow 0$ as $n\rightarrow\infty$, the assumption $H(g(X_1^n,X_2^n))\leq H(f(X_1^n,X_2^n))$ as $n\rightarrow\infty$ gives  the upper bound
\begin{align}
R_1+R_2\leq \max_{p(x_1)p(x_2)}I(Y_1;X_1,X_2).
\label{eq:upper}
\end{align}

Under our assumption in the theorem that the sequence of codes $\Cc_1^{(n)}, \Cc_2^{(n)}$ are \textit{good} computation codes for the channel $W_1$, we know that  the function $f(X_1^n,X_2^n)$ can be decoded reliably over $W_1$, and furthermore we have the achievable computation sum-rate
\begin{align}
R_1+R_2=\max_{p(x_1)p(x_2)}I(Y_1;X_1,X_2)+\delta
\label{eq:good_computation}
\end{align}
for some $\delta> 0$, by Definition \ref{def:good_computation_codes}. However, this implies immediately that the pair $(X_1^n,X_2^n)$ can not be decoded reliably over $W_2$. Indeed, if the decoder of $W_2$  could decode both codewords,  the achievable sum-rate in (\ref{eq:good_computation}) directly contradicts the upper bound in (\ref{eq:upper}). This proves that this sequence of codes cannot be used as multiple access codes for the channel $W_2$. 
\end{IEEEproof}

\begin{IEEEproof}[Proof of Theorem \ref{thm:duality_2}]
Again we consider two multiple access channels $W_1, W_2$ as described in the theorem. We assume temporarily that the pair of codes $(\mathcal C_1^{(n)}, \mathcal C_2^{(n)})$ are used for computation over $W_1$, and used for multiple access over $W_2$.  Under the assumption in the theorem, both codewords $(X_1^n,X_2^n)$ can be recovered over the channel  $W_2$ with the rate pair $(R_1,R_2)$. Suppose the function $f(X_1^n,X_2^n)$ can also be reliably decoded over the channel $W_1$, then it must satisfy
\begin{align*}
R_1+R_2\leq \max_{p(x_1)p(x_2)}I(X_1, X_2; Y_1)
\end{align*}
as shown in the upper bound (\ref{eq:upper}).  By Definition \ref{def:good_computation_codes}, this sequence of codes are not \textit{good computation codes} for the channel $W_1$, since the sum-rate is not larger than the  sum capacity of $W_1$.
\end{IEEEproof}

\section{Distributed Source Coding and Computation Duality}\label{sec:dsc-duality}
\begin{figure}[!hbt]
\begin{center}
\footnotesize
\psfrag{e1}[c]{Encoder 1}
\psfrag{e2}[c]{Encoder 2}
\psfrag{d1}[c]{Decoder 1}
\psfrag{d2}[c]{Decoder 2}
\psfrag{m1}[c]{$M_1$}
\psfrag{m2}[c]{$M_2$}
\psfrag{x1}[c]{$X^n_1$}
\psfrag{x2}[c]{$X^n_2$}
\psfrag{y1}[c]{$Y^n_1$}
\psfrag{y2}[c]{$Y^n_2$}
\psfrag{z1}[c]{$X_1^n+X^n_2$}
\psfrag{z2}[c]{$\qquad\quad (X_1^n, X_2^n)$}
\includegraphics[scale=1]{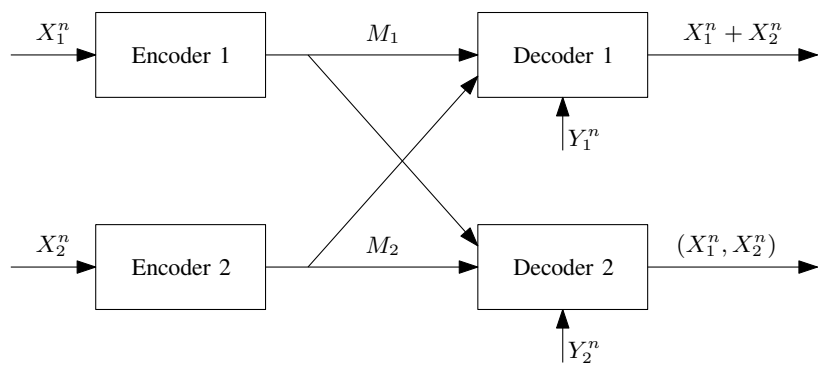}
\caption{Two-sender two-receiver distributed source coding network.}
\label{fig:twoSW}
\end{center}
\end{figure}

In this section, we establish duality results for  distributed source coding.  Consider the two-sender two-receiver distributed source coding network in Figure \ref{fig:twoSW}. Two correlated sources $X_1^n, X_2^n$ are encoded by Encoder $1$ and Encoder $2$, respectively. Decoder $1$ wishes to decoder a function of the sources $f(X_1^n, X_2^n)$ with side information $Y_1^n$, and decoder $2$ wishes to decode the two sources with side information $Y_2^n$. We will show that good computation codes cannot be used for distributed source coding, and vice versa.

To state the problem formally, consider a discrete memoryless source (DMS) triple $(X_1, X_2, Y)$ that consists of three finite alphabets $\mathcal X_1,\mathcal X_2,\mathcal Y$ and a joint pmf of the form
\begin{align}\label{eq:source-pmf}
p(x_1, x_2, y)=p(x_1, x_2)W(y|x_1, x_2).
\end{align}
A formal definition of a distributed source coding (DSC) code is given as follows.

\begin{definition}[Distributed Source Coding Codes]\label{def:dsc_code}
A $(2^{nR_1}, 2^{nR_2}, n)$ code for distributed source coding consists of 
\begin{itemize}
\item two encoders, where encoder $k=1,2,$ assigns an index $m_1(x_1^n)\in[2^{nR_1}]$ to each sequence $x^n_1\in\Xc^n_1$, and
\item a decoder that assigns an estimate $(\xh_1^n, \xh_2^n)$ to each index pair $(m_1, m_2)\in[2^{nR_1}]\times[2^{nR_2}]$ and side information $y^n\in \Yc^n$.
\end{itemize}
\end{definition}

The probability of error for a distributed source code is defined as 
\begin{align}
P_\e^{(n)} = \P\{(\Xh_1^n, \Xh_2^n)\neq (X^n_1, X^n_2)\}.
\end{align}

A rate pair $(R_1, R_2)$ is said to be achievable for distributed source coding if there exists a sequence of $(2^{nR_1}, 2^{nR_2}, n)$ codes such that $\lim_{n\to\infty}P_\e^{(n)}=0$. The Slepian--Wolf (SW) region, given below\footnote{Including a slight modification accounting for the side information at Decoder 2.}, gives a complete characterization of the achievable rate region:
\begin{align}\label{eq:SW}
R_1 &> H(X_1|X_2, Y_2), \nonumber\\
R_2 &> H(X_2|X_1, Y_2), \nonumber\\
R_1+R_2 &> H(X_1, X_2| Y_2).
\end{align}

The following definition formalizes the concept of computation codes for distributed source coding used in this paper.
\begin{definition}[Computation codes for DSC]\label{def:dsc_comp_code}
A $(2^{nR_1}, 2^{nR_2}, n, f)$ {\em computation code} for a DMS triple $(X_1, X_2, Y)$ consists of two message sets and two encoders defined as in Definition~\ref{def:dsc_code} and
\begin{itemize}
\item a function $f:\mathcal X_1\times\mathcal X_2\mapsto \mathcal F$ for some image $\mathcal F$, and
\item a decoder that maps an estimated function value $\fh^n\in\mathcal F^n$ to each index pair $(m_1, m_2)\in[2^{nR_1}]\times[2^{nR_2}]$ and side information $y^n\in \Yc^n$.
\end{itemize}
The average probability of error for computation is defined as
\begin{align}\label{eq:pe-cp-dsc}
P_\e^{(n)}=\P\{F^n\neq \hat{F}^n\},
\end{align}
where $F^n=f(X_{11}, X_{21}), \ldots, f(X_{1n}, X_{2n})$ denotes an element-wise application of the function $f$ on the pair $(X_1^n, X_2^n)$. We say a rate pair $(R_1, R_2)$ is {\em achievable for computation} if there exists a sequence of $(2^{nR_1}, 2^{nR_2}, n)$ computation codes such that $\lim_{n\to\infty}P_\e^{(n)}=0$.
\end{definition}

Since the function $f(X_1^n, X_2^n)$ can be computed directly at the receiver if the individual codewords $(X_1^n, X_2^n)$ are known, a $(2^{nR_1}, 2^{nR_2}, n)$ DSC code is readily a $(2^{nR_1}, 2^{nR_2}, n, f)$ computation code for \textit{any} function $f$. More interesting are the computation codes with rates outside the optimal DSC rate region, i.e. $(R_1,R_2)\notin \Rs_\text{SW}$. We refer to such codes as {\em good} computation codes for DSC.

\begin{definition}[Good computation codes]\label{def:good_computation_codes_DSC}
Consider a sequence of $(2^{nR_1}, 2^{nR_2},n,f)$-computation codes for a DMS $(X_1, X_2, Y)\sim p(x_1, x_2)W(y|x_1,x_2)$. We say that the computation codes are {\em good computation codes}  for this DMS,  if $(R_1,R_2)$ is achievable for computation, and 
\begin{align*}
R_1+R_2< H(X_1, X_2|Y).
\end{align*}
Namely the sum-rate of the two codes is smaller than the sum-rate constraint in $\Rs_\text{SW}$.
\end{definition}

When $(X_1, X_2, Y_1)$ are from a finite field and the function to compute is the sum $X_1+X_2$, the stand-alone problem associated with Decoder 1 (i.e. without Decoder 2) was considered in the seminal work of K\"orner and Marton~\cite{Korner--Marton1979}. In their work, it was shown that using the same linear code at both encoders, a rate pair $(R_1, R_2)$ is achievable for computing the sum of the sources if\footnote{Including a slight modification accounting for the side information at Decoder 1.}
\begin{align}\label{eq:KM}
R_1&> H(X_1+X_2|Y_1),\nonumber\\
R_2&> H(X_1+X_2|Y_1).
\end{align} 
We denote this rate region by $\mathscr{R}_\text{KM}$.

\begin{definition}[$g$-SI]
Consider a DMS $(X_1,X_2,Y)$ described by $p(x_1,x_2)W(y|x_1,x_2)$. We say $Y$ is a \textit{$g$-side information ($g$-SI)} if there is a function $g:\mathcal X_1\times \mathcal X_2\mapsto \mathcal F$ for some set $\mathcal F$ such that the following Markov chain holds
\begin{align*}
Y\to g(X_1,X_2)\to (X_1, X_2).
\end{align*}
\label{def:gSI}
\end{definition}

\subsection{Main Results}
In this subsection we show that for any sequence of codes, there is an intrinsic tension between their capability for computation and their capability for distributed source coding. 

\begin{theorem}[DSC-Computation Duality 1]\label{thm:dsc-duality1}
Consider a two-sender two-receiver DSC network in Fig.~\ref{fig:twoSW}. Assume that the DMS $(X_1, X_2, Y_1, Y_2)\sim p(x_1, x_2, y_1, y_2)$ is given by $p(x_1,x_2)W_1(y_1|x_1,x_2)W_2(y_2|x_1,x_2)$  where the side information $Y_2$ is a $g$-SI according to Definition \ref{def:gSI}. Further assume that a sequence of codes $(\Cc_1^{(n)},\Cc_2^{(n)})$ is good for computing the function $f$ with the side information $Y_1$, namely the sum rate of the codes satisfies $R_1+R_2 < H(X_1, X_2|Y_1)$. Then this sequence of codes cannot be used as DSC codes with side information $Y_2$ (i.e., the receiver cannot recover both sources correctly), if it holds that
\begin{align}
H(g(X_1^n,X_2^n)|M_1,M_2,Y_1^n)\leq H(f(X_1^n,X_2^n)|M_1,M_2,Y_1^n) \text{ as } n\rightarrow\infty.
\end{align}
\end{theorem}

\begin{remark}
More precisely, we will show in the proof that the optimal rate region of the two-sender two-receiver network is bounded by 
\begin{align}
R_1+R_2 \ge H(X_1, X_2|Y_1). \label{eq:sumrate_bnd2}
\end{align}
Notice that the side information $Y_2$ does not directly appear in the above inequality.
\end{remark}

\begin{remark}
Same as in Theorem \ref{thm:duality_1}, the entropy inequality in the above theorem is in general difficult to verify. Nevertheless an interesting special case is when  $g=f$ where this condition is trivially satisfied (see Example~\ref{ex:dsc1}).
\end{remark}

The following theorem gives a complementary result.

\begin{theorem}[DSC-Computation Duality 2]\label{thm:dsc-duality2}
Consider a two-sender two-receiver distributed source coding network in Fig.~\ref{fig:twoSW}. Assume that the DMS $(X_1, X_2, Y_1, Y_2)\sim p(x_1, x_2, y_1, y_2)$ is given by $p(x_1,x_2)W_1(y_1|x_1,x_2)W_2(y_2|x_1,x_2)$  where the side information $Y_2$ is a $g$-SI according to Definition \ref{def:gSI}. Further assume that a sequence of codes $(\Cc_1^{(n)},\Cc_2^{(n)})$ is a sequence of DSC codes for side information $Y_2$. Then, this sequence of codes cannot be good computation codes for computing the function $f$ with side information $Y_1$, if it holds that
\begin{align}
H(g(X_1^n,X_2^n)|M_1,M_2,Y_1^n)\leq H(f(X_1^n,X_2^n)|M_1,M_2,Y_1^n) \text{ as } n\rightarrow\infty.
\label{condition:source_coding}
\end{align}
\end{theorem}



Before presenting the proofs, we  give a few examples to illustrate the results. 
\subsection{Examples}

\begin{example}\label{ex:dsc1}
Consider the case in which $Y_1=\emptyset$, $Y_2=X_1\oplus X_2$, and $(X_1, X_2)$ are doubly symmetric binary sources, i.e., $\P\{(X_1, X_2)=(1, 1)\}=\P\{(X_1, X_2)=(0, 0)\}=(1-p)/2$, $\P\{(X_1, X_2)=(0, 1)\}=\P\{(X_1, X_2)=(1, 0)\}=p/2$ for some $p\in[0,1/2]$. It can be checked directly that the conditions in Theorem \ref{thm:dsc-duality1} are satisfied for this setup, and we have the promised duality results.

\begin{figure*}[h!]
\begin{center}
\footnotesize
\psfrag{r1}[c]{$R_1$}
\psfrag{r2}[c]{$R_2$}
\psfrag{x1}[r]{$H(X_1|Y_2)$}
\psfrag{x2}[c]{$H(X_2|Y_2)$}
\psfrag{h1}[r]{$H_2(p)$}
\psfrag{h2}[c]{$H_2(p)$}
\psfrag{a1}[c]{$D$}
\psfrag{b1}[c]{$E$}
\psfrag{c1}[c]{$A$}
\psfrag{d1}[c]{}
\psfrag{e1}[c]{}
\psfrag{f1}[c]{$B$}
\psfrag{g1}[c]{$C$}
\includegraphics[scale=1]{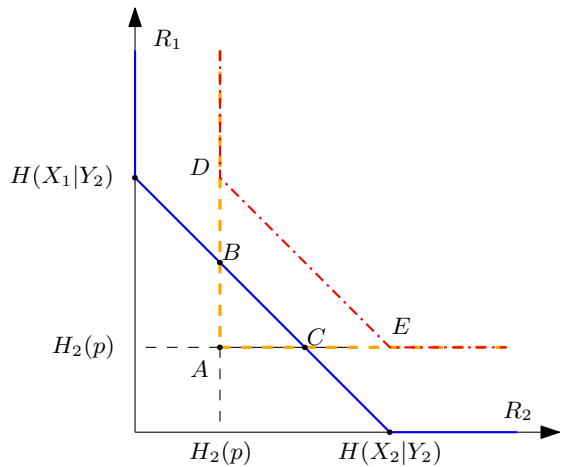}
\end{center}
\caption{Rate regions $\mathscr{R}_\text{KM}$ and $\mathscr{R}_\text{SW}$ in Example~\ref{ex:dsc1}. The solid line rate region is $\mathscr{R}_\text{SW}$ for Decoder 2, the dashed rate region and the dashed-dotted rate region is $\mathscr{R}_\text{KM}$ and $\mathscr{R}_\text{SW}$ for Decoder 1, respectively. Here, $H_2(p)$ is the binary entropy function.}\label{fig:region1}
\end{figure*}
\end{example}

The rate regions $\mathscr{R}_\text{KM}$ and $\mathscr{R}_\text{SW}$ for this example are given in Fig.~\ref{fig:region1}.  The duality results have some interesting implictaions on random linear codes. Wyner  \cite{Wyner1974b} has shown  that nested linear codes can achieve the corner points of $\mathscr{R}_\text{SW}$. The proof technique in~\cite{Wyner1974b} relies on first recovering one of the sources, say $X_1$, with optimal linear source coding (with $Y_2$ side information) then sequentially recovering $X_2$ by treating $X_1$ as additional side information which achieves the corner point $(H(X_1|Y_2), H(X_2|X_1, Y_2))$. By time sharing with the other corner point attained by switching the decoding order, the whole $\mathscr{R}_\text{SW}$ region is achievable via nested linear codes. An interesting question is whether it is possible to attain the whole region $\mathscr{R}_\text{SW}$ using optimal joint decoding (without time sharing). 

Theorem~\ref{thm:dsc-duality1} implies  that while the corner points are achievable with nested linear codes, it is impossible to attain the whole $\mathscr{R}_\text{SW}$ rate region even under optimal decoding (maximum likelihood decoders). To see this, consider the setting in Example~\ref{ex:dsc1}. The important observation is that while nested linear codes can be used to attain the corner points of $\mathscr{R}_\text{SW}$, they are also good computation codes~\cite{Korner--Marton1979}. In this example, the codes whose rate pairs lie inside the boundary region $A-E-D$ in Figure~\ref{fig:region1} are good computation codes. Hence according to Theorem \ref{thm:dsc-duality1}, these nested linear codes with such rate pairs cannot be used for distributed source coding under side information $Y_2$, i.e., the points inside the boundary region $B-C-E-D$ are  not achievable using these codes.



\begin{example}
Consider a discrete memoryless binary source pair $X_1,X_2\in\{0,1\}$. If two codes $\mathcal C_{1}^{(n)},\mathcal C_{2}^{(n)}$ are good for computing $x_1^n+x_2^n$ with any side information $Y_1^n$ (addition performed element-wise in $\mathbb R$), then this sequence of codes can not be used as distributed source coding (DSC) codes  with side information $x_1^n\cdot x_2^n$.

The result holds according to Theorem \ref{thm:dsc-duality1} by choosing
\begin{align*}
f(x_1,x_2)&= x_1+x_2\\
g(x_1,x_2)& = x_1\cdot x_2.
\end{align*}
The condition (\ref{condition:source_coding}) is fulfilled since
\begin{align*}
H(X_1^n + X_2^n|M_1,M_2,Y_1) &= H(X_1^n + X_2^n, X_1^n
\cdot X_2^n|M_1,M_2,Y_1)\\
&\geq H(X_1^n
\cdot X_2^n|M_1,M_2,Y_1)
\end{align*}
\end{example}

\begin{example}
Consider a discrete memoryless binary source pair $X_1,X_2\in\{0,1\}$. If two codes $\mathcal C_{1}^{(n)},\mathcal C_{2}^{(n)}$ are good for computing $x_1^n\cdot x_2^n$ with any side information $Y_1^n$ (multiplication performed element-wise in $\mathbb R$), then this sequence of codes can not be used as DSC codes  with side information $Y_2^n = x_1^n\cdot x_2^n+Z^n$ where $Z^n\in\{0,1\}^n$ is a i.i.d. random binary sequence independent of $Y_1^n$ and the sources.

The result holds according to Theorem \ref{thm:dsc-duality1} by choosing
\begin{align*}
f(x_1,x_2)=g(x_1,x_2)= x_1\cdot x_2
\end{align*}
and noticing that $Y_2^n$ is a $g$-SI.
\end{example}

\subsection{Proofs}
\begin{IEEEproof}[Proof of Theorem~\ref{thm:dsc-duality1}]
We assume temporarily that the pair of codes $(\mathcal C_1^{(n)}, \mathcal C_2^{(n)})$ are used for computation at Decoder 1, and used for distributed lossy source coding at Decoder 2. In other words, the function $f(X_1^n,X_2^n)$ can be  decoded  reliably using $Y_1^n$, and the pair $(X_1^n,X_2^n)$ can be recovered reliably using $Y_2^n$.  Then,
\begin{align*}
n(R_1+R_2) &= H(M_1, M_2) \\
&= H(M_1, M_2)+H(X^n_1, X_2^n | M_1, M_2, Y_2^n)-H(X^n_1, X_2^n | M_1, M_2, Y_2^n)\\
&\stackrel{(a)}{\ge} H(M_1, M_2)+H(X^n_1, X_2^n | M_1, M_2, Y_2^n)-n\e_n\\
&= H(M_1, M_2)+H(X^n_1, X_2^n, g(X_1^n,X_2^n) | M_1, M_2, Y_2^n)-n\e_n\\
&\ge H(M_1, M_2)+H(X^n_1, X_2^n | M_1, M_2, Y_2^n, g(X_1^n,X_2^n))-n\e_n\\
&\stackrel{(b)}{=} H(M_1, M_2)+H(X^n_1, X_2^n | M_1, M_2, g(X_1^n,X_2^n))-n\e_n\\
&\stackrel{(c)}{\geq} H(M_1, M_2)+H(X^n_1, X_2^n | M_1, M_2, g(X_1^n,X_2^n),Y_1^n)\\
 &\quad +H(g(X_1^n,X_2^n)|M_1,M_2,Y_1^n)-H(f(X_1^n,X_2^n)|M_1,M_2,Y_1^n)-n\e_n\\
&\stackrel{(d)}{\ge} H(M_1, M_2)+H(X^n_1, X_2^n | M_1, M_2, g(X_1^n,X_2^n))+H(g(X_1^n,X_2^n)|M_1, M_2,Y_1^n)-2n\e_n\\
&\geq H(M_1,M_2|Y_1^n)+H(X_1^n,X_2^n,g(X_1^n,X_2^n)|M_1,M_2,Y_1^n)-2n\epsilon_n\\
&= H(M_1,M_2,X_1^n,X_2^n,g(X_1^n,X_2^n)|Y_1^n)-2n\epsilon_n\\
&= H(X_1^n,X_2^n|Y_1^n)-2n\e_n\\
& = \sum_{i=1}^n H(X_{1i}, X_{2i}|Y_{1i})-2n\e_n,
\end{align*}
where  step $(a)$ and $(d)$ is from Fano's inequality, step $(b)$ follows from the Markovity $Y^n_2\to g(X_1^n,X_2^n) \to (X^n_1, X_2^n)$, and step $(c)$ follows from our assumption in the theorem. 
Overall, as $n\rightarrow\infty$ we have the lower bound
\begin{align}\label{eq:dualityLB}
R_1+R_2 \ge H(X_1, X_2|Y_1). 
\end{align}
Under assumption that $(\Cc_1^{(n)}, \Cc_2^{(n)})$ are good computation codes with side information $Y_1$, it satisfies the condition $R_1+R_2 < H(X_1, X_2|Y_1)$. However, this directly contradicts the lower bound~\eqref{eq:dualityLB}. This proves that this sequence of codes cannot be used as distributed source coding codes with side information $Y_2$.
\end{IEEEproof}

\begin{IEEEproof}[Proof of Theorem~\ref{thm:dsc-duality2}]
Consider the network in Figure \ref{fig:twoSW} again. We assume temporarily that the pair of codes $(\mathcal C_1^{(n)}, \mathcal C_2^{(n)})$ are used for computation with side information $Y_1$, and used for distributed source coding with side information $Y_2$.  Under the assumption in the theorem, both sources $(X_1^n,X_2^n)$ can be recovered with side information $Y_2$ with rates $(R_1,R_2)$. Suppose the function $f(X_1^n,X_2^n)$ can also be reliably recovered with side information $Y_1$, then it must satisfy
\begin{align*}
R_1+R_2\geq  H(X_1, X_2|Y_1),
\end{align*}
as shown in the lower bound (\ref{eq:dualityLB}).  By Definition \ref{def:good_computation_codes}, this sequence of codes are not \textit{good computation codes} with the side information $Y_1$, since the sum-rate is not smaller than the  sum-rate in the Slepian--Wolf rate region under side information $Y_1$.
\end{IEEEproof}

\section{Duality over the Gaussian MAC}
\label{sec:GMAC}
Computation codes, primarily lattice codes have been studied intensively in Gaussian multiple access channels. In this section we specialize the results in previous sections to additive channel models, and focus on decoding the sum of two codewords. In particular we consider the symmetric Gaussian MAC 
\begin{align}
Y^n=x^n_1+x^n_2+Z^n
\label{eq:MAC}
\end{align}
where $Z_i\sim \mathcal N(0,N)$, $i\in[n]$ is an additive white Gaussian noise. Both channel inputs have the same average power constraint $\sum_{i=1}^n x^2_{ki}\le nP$,  $k=1,2$. For the sake of notation, we will define the signal-to-noise ratio (SNR) to be $\snr:=P/N$, and denote such a symmetric two-user Gaussian MAC  as GMAC($\snr$).

\subsection{Computing sums of codewords}
A Gaussian MAC naturally adds two codewords through the channel, hence it is particularly beneficial for the decoder to decode the sum of the two codewords. In this section, we will only focus on decoding the sum of the codewords, i.e., the function $f$ is defined to be $f=x_1+x_2$. For this channel, it is well known that nested lattice codes are good computation codes for computing the sum $x_1^n+x_2^n$. In particular, it is shown in  \cite{NazerGastpar_2011} that nested linear codes is able to achieve  a computation rate pair $(R_1,R_2)$ if it satisfies 
\begin{align*}
R_1&< \sfrac{1}{2}\log \left(\sfrac{1}{2}+\snr\right),\\
R_2&< \sfrac{1}{2}\log \left(\sfrac{1}{2}+\snr\right).
\end{align*}
It is easy to see that the sum-rate $R_1+R_2$ is outside the capacity region of the Gaussian MAC if $\snr>3/2$.

Now we come back to the system depicted in Figure \ref{fig:compoundMAC}. 
Specializing the duality results to this  scenario, we have the following theorems.
\begin{corollary}[Gaussian MAC duality 1]
Let $\snr_1$ be a fixed but arbitrary value  and  let $(\mathcal C_1^{(n)},\mathcal C_2^{(n)})$ be a sequence of good computation codes for GMAC$(\snr_1)$. Then this sequence of good computation codes cannot be a sequence of  multiple access codes for GMAC$(\snr_2)$, for \textit{any} $\snr_2$.
\label{thm:duality_1_GMAC}
\end{corollary}
\smallskip

\begin{figure}[htb!]
\begin{center}
\begin{tikzpicture}[every path/.append style={thick}]
\begin{axis}[
	xmin=0, xmax=1.4,
	ymin=0, ymax=1.4,
	xticklabels={}, yticklabels={}, ticks=none,
	xlabel={$R_1$}, ylabel={$R_2$},
	axis lines=middle,
	width=7cm,
	axis equal image,
	axis on top=true,
	clip=false,
	]
    \coordinate (origin) at (0,0);
	\coordinate (A) at (.6,1);
	\coordinate (B) at (1,.6);
	\coordinate (C) at (.9,1.2);
	\coordinate (D) at (1.2,.9);
	\coordinate (AA) at (.9,.9);
	\draw[blue] (origin)-|(B)--(A)-|(origin);
	\draw[red] (origin)-|(D)--(C)-|(origin);
	\draw[dashed]
    	(AA -| origin)--(AA)--(AA |- origin);
	\node[draw,fill=white,shape=circle,inner sep=1.2pt,label={[label distance={-.5ex}]60:\textsf{A}}] at (AA) {};
    \draw[dotted] (origin)--(1.1,1.1);
\end{axis}
\end{tikzpicture}
\end{center}
\caption{MAC capacity regions for $\snr_1$ (blue), $\snr_2$ (red) where $\snr_2>\snr1$. The point ${\sf A}$ is achievable by a pair of good computation codes $\Cc_1,\Cc_2$ over GMAC($\snr_1$) with rate $(R_1,R_2)$. While the rate pair $(R_1,R_2)$ is included in the capacity region of GMAC($\snr_2$), the codes $\Cc_1,\Cc_2$ cannot be used as multiple access codes for GMAC($\snr_2$).}
\label{fig:two_MACs}
\end{figure}
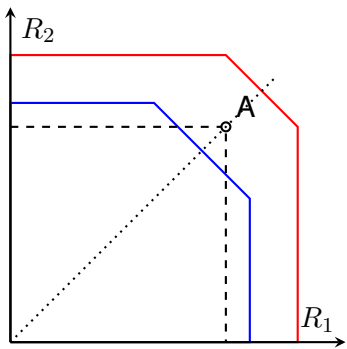

\begin{remark}
By the definition of good computation codes for GMAC($\snr_1$), the rate pair $(R_1,R_2)$ is outside the capacity region $\mathcal C_{mac}(\snr_1)$. Hence it obviously cannot be an achievable rate pair for multiple access in a Gaussian MAC with an SNR value smaller or equal to $\snr_1$.
The question of interest is if this sequence of good computation codes  can be used for multiple access over a Gaussian MAC when its SNR is much larger than $\snr_1$.
The above result shows that good computation codes cannot be used for multiple access even with an arbitrarily large SNR. Figure \ref{fig:two_MACs} gives an illustration of this result.
\end{remark}

\begin{proof}
The function to be computed is defined to be $f(x_1,x_2)=x_1+x_2$. Also notice that a GMAC($\snr$) also a $f$-MAC for any $\snr$. Hence the condition (\ref{eq:condition}) holds  and our claim follows directly from Theorem \ref{thm:duality_1}.
\end{proof}

The following theorem gives a complementary result, whose proof follows that in Theorem \ref{thm:duality_2}.
\begin{corollary}[Gaussian MAC duality 2]
Let $\snr_2$ be a fixed but arbitrary value, and let $(\Cc_1^{(n)}, \Cc_2^{(n)})$ be a sequence of $(2^{nR_1},2^{nR_2},n)$ multiple access codes for the GMAC($\snr_2$) with arbitrary $R_1,R_2$. Then this sequence of codes cannot be  good computation codes for the GMAC($\snr_1$), for \textit{any} $\snr_1$.
\label{thm:duality_2_GMAC}
\end{corollary}

\subsection{Sensitivity to channel coefficients}
For decoding the sum of codewords $X_1^n+X_2^n$, the duality results for Gaussian MAC depend crucially on the channel gains. Theorem \ref{thm:duality_1_GMAC} and \ref{thm:duality_2_GMAC} are established for the case when the channel gains are matched to the coefficients in the sum (namely $(1,1)$). Now we show that if the channel gains and the coefficients in the sum  are not matched, duality results do not hold in general.

\begin{proposition}
There exists a sequence of codes $(\mathcal C_1^{(n)},\mathcal C_2^{(n)})$ such that they are good computation codes for the GMAC($1,1,N_1$), and they can also be used for multiple access over the GMAC($1,c,N_2$) for any integer $c\geq 2$, for some $N_1, N_2> 0$.
\end{proposition}

\begin{proof}
It is shown in \cite{lim_joint_2016}  that for a GMAC($1,c,N_2$), there exists a sequence of nested linear codes $\mathcal C_1^{(n)}, \mathcal C^{(n)}$ that can be used for multiple access for this channel if the rate pair $(R_1,R_2)$ satisfies
\begin{align}
R_1 &< \C(P_1/N_2)\nonumber\\
R_2 &< \C(P_2/N_2)\nonumber\\
R_1+R_2 &< \C((1+c^2)P/N_2)\nonumber\\
R_1 &< \max_{\mathbf{a} \in \mathbb{Z}^2 \setminus \{\mathbf{0}\}}\min\{I_{\mathsf{CF},1}(\mathbf{a}),  \C((1+c^2)P/N_2)-I_{\mathsf{CF},2}(\mathbf{a})\} \text{ or } \label{eq:lmac1}\\
R_2 &< \max_{\mathbf{a} \in \mathbb{Z}^2 \setminus \{\mathbf{0}\}}\min\{I_{\mathsf{CF},2}(\mathbf{a}),  \C((1+c^2)P/N_2)-I_{\mathsf{CF},1}(\mathbf{a})\} \label{eq:lmac2}
\end{align}
where $\C(x)=\sfrac{1}{2}\log(1+x)$,
\begin{align}
	I_{\mathsf{CF},1}(\mathbf{a}) &= \sfrac{1}{2}\log\left(\frac{1+(1+c^2)P}{(a_1c-a_2 )^2P+a_1^2+a_2^2}\right)+\log\gcd(\mathbf{a}), \label{eq:icf1}\\
	I_{\mathsf{CF},2}(\mathbf{a}) &= \sfrac{1}{2}\log\left(\frac{1+(1+c^2)P}{(a_1c-a_2 )^2P+a_1^2+a_2^2}\right)+\log\gcd(\mathbf{a}), \label{eq:icf2}
\end{align} 
The above rate region is denoted by $\Rc_\text{LMAC}$. For simplicity, we define an inner bound $\tilde\Rc_\text{LMAC}$ on $\Rc_\text{LMAC}$ by choosing $\mathbf{a}=[1,\, c]$ in the maximization of~\eqref{eq:lmac1} and~\eqref{eq:lmac2}.
Moreover, we let
\begin{align}\label{eq:condpn2}
\frac{P}{N_2} > \frac{(1+c^2)^2-1}{(1+c^2)},
\end{align}
such that~\eqref{eq:lmac1} and~\eqref{eq:lmac2} are simplified by
\begin{align}
R_k<\sfrac{1}{2}\log\left(1+c^2\right),\quad k=1,2.
\end{align}

It is also shown in~\cite{lim_joint_2016} that nested lattice codes can be used to compute the sum of two codewords for the GMAC$(1,1,N_1)$ if the rate pair is included in the rate region:
\begin{align*}
\mathcal R_{CF}:=\{(R_1,R_2)| R_1&<\frac{1}{2}\log(1/2+P/N_1)\\
R_2&<\sfrac{1}{2}\log(1/2+P/N_1)\}.
\end{align*}
If we require that a pair $(R_1,R_2)\in \mathcal R_{CF}$ to be a rate pair of good computation codes, it should satisfy that
\begin{align*}
2 \cdot\sfrac{1}{2}\log(1/2+P/N_1)>\sfrac{1}{2}\log(1+2P/N_1)
\end{align*}
which imposes the constraint $P/N_1>3/2$, which we will always assume in the proof.

Now we will show that for some $N_1, N_2$, we can find a rate pair $(R_1^*,R_2^*)$ which lies in both $\tilde{\mathcal R}_{LMAC}$ and $\mathcal R_{CF}$. This shows that a pair of nested linear codes with this rate pair can be used for multiple access for GMAC$(1,c,N_2)$ and used for computation for GMAC$(1,1,N_1)$.  Specifically, we  choose the rate to be 
\begin{align*}
R_1^*=R_2^*=\sfrac{1}{2}\log 3-\epsilon
\end{align*}
for some small $\epsilon>0$. We  first show that $(R_1^*,R_2^*)\in\tilde{\Rc}_{LMAC}$ for some $N_2$. 



By choosing $c>\sqrt{2}$, the RHS terms of the last two constraints in $\tilde{\Rc}_{LMAC}$ is larger or equal to $\frac{1}{2}\log 3$. Furthermore, it is easy to see that if we choose $N_2$ such that $P/N_2>2$ and $(1+c)^2P/N_2\geq 8$ are satisfied (along with the previous assumption~\eqref{eq:condpn2}), the rate pair $R_1^*, R_2^*$ is included in $\tilde{\Rc}_{LMAC}$.

To show that $(R_1^*,R_2^*)\in\mathcal R_{CF}$ for some $N_1$,  we only need to choose $N_1$ such that  $1/2+P/N_1> 3$, or equivalently $P/N_1> 5/2$ (notice it satisfies the constraint $P/N_1>3/2$).  This completes the claim.
\end{proof}

\begin{remark}
Combining this result and Theorem \ref{thm:duality_1},  we can conclude that for the considered nested linear codes  with rate $(R_1^*,R_2^*)$ in the proof, we have $H(X_1^n+cX_2^n)>H(X_1^n+X_2^n)$ for any integer $c\geq 2$, where $X_1^n, X_2^n$ denote the uniformly chosen random codewords.
\end{remark}

\section{Discussions}\label{sec:discussions}

Computation codes have been studied in many network information theory problems. The main motivation behind the use of such codes is that it is more efficient to compute a function of codewords than to recover the individual codewords separately. However, this efficiency comes with a cost.
In this work we characterized duality relations between computation codes and codes for multiple-access and distributed compression.
Our results are not limited to a specific computation code such as lattice or nested linear codes and apply to any efficient computation code.
We showed that if the multiple access channel is ``aligned'' with the target computation  function, then good computation codes must possess certain structure such that the individual messages cannot be recovered at the destination node, regardless of what decoder is used. We further explored a source coding setting to characterize a similar relationship. If the side information is ``aligned'' with the target  function to be computed, then good computation codes must possess certain structure such that the individual sources cannot be recovered at the destination node. 

%
%
%
%
%

\section*{Acknowledgment}
The work of Jingge Zhu was supported by the Swiss National Science Foundation under Project P2ELP2\_165137 and the work of Sung Hoon Lim was supported
in part by the National Research Foundation of Korea (NRF) Grant NRF-2017R1C1B1004192.


\bibliographystyle{IEEEtran}
\bibliography{duality}

\end{document}